\documentclass[sigconf]{acmart}

\usepackage{booktabs} 

\copyrightyear{2019}
\acmYear{2019} 
\setcopyright{iw3c2w3}
\acmConference[WWW '19 Companion]{Companion Proceedings of the 2019 World Wide Web Conference}{May 13--17, 2019}{San Francisco, CA, USA}
\acmBooktitle{Companion Proceedings of the 2019 World Wide Web Conference (WWW '19 Companion), May 13--17, 2019, San Francisco, CA, USA}
\acmPrice{}
\acmDOI{10.1145/3308560.3316751}
\acmISBN{978-1-4503-6675-5/19/05}

\usepackage{balance}

\begin{document}

%
\title{Detecting and Gauging Impact on Wikipedia Page Views}

%

\author{Xiaoxi Chelsy Xie}
\email{cxie@wikimedia.org}
\affiliation{\institution{Wikimedia Foundation}}

\author{Isaac Johnson}
\email{isaac@wikimedia.org}
\affiliation{\institution{Wikimedia Foundation}}

\author{Anne Gomez}
\email{agomez@wikimedia.org}
\affiliation{\institution{Wikimedia Foundation}}

%
\renewcommand{\shortauthors}{Xie et al.}

%
\begin{abstract}
Understanding how various external campaigns or events affect readership on Wikipedia is important to efforts aimed at improving awareness and access to its content. In this paper, we consider how to build time-series models aimed at predicting page views on Wikipedia with the goal of detecting whether there are significant changes to the existing trends. We test these models on two different events: a video campaign aimed at increasing awareness of Hindi Wikipedia in India and the page preview feature roll-out---a means of accessing Wikipedia content without actually visiting the pages---on English and German Wikipedia. Our models effectively estimate the impact of page preview roll-out, but do not detect a significant change following the video campaign in India. We also discuss the utility of other geographies or language editions for predicting page views from a given area on a given language edition.
\end{abstract}

%
%


%

%
\maketitle
\section{Introduction}
Wikipedia is the fifth-most-visited website worldwide \cite{alexa_wikipedia.org_2018} at 190 billion page views in 2018 alone \cite{erhart_wikipedias_2019} and is turned to by readers for all sorts of reasons ranging from simple curiosity to fact-checking to making a personal decision \cite{singer_why_2017}. Despite its success, there are still many regions in the world where it is relatively unknown \cite{gill_how_2018} or access is blocked \cite{clark_analyzing_2017}. In an attempt to improve access and awareness worldwide, the Wikimedia Foundation (WMF) has conducted various campaigns and efforts aimed at improving access to Wikipedia in various regions.\footnote{https://meta.wikimedia.org/wiki/New\_Readers}

Many researchers have sought to estimate the impact of external events on Wikipedia page views. This has included the effect of posting Wikipedia articles on external websites \cite{vincent_examining_2018,moyer_determining_2015}, privacy concerns on viewing of sensitive Wikipedia articles \cite{penney_chilling_2016}, and censorship \cite{zhang_group_2011}. Conversely, much research has also sought to use Wikipedia page views as a predictor---i.e. \emph{nowcasting} or \emph{forecasting} \cite{priedhorsky_measuring_2017}---of external entities such as the stock market \cite{moat_quantifying_2013}, box office returns for movies \cite{mestyan_early_2013}, and disease incidence \cite{priedhorsky_measuring_2017}.

Evaluating the impact, however, of a given campaign or external event can be difficult. Daily page views to Wikipedia projects can be quite noisy, being affected by weekly, seasonal, holiday-related trends \cite{ten_thij_modeling_2012}. A change in the volume of page views following an awareness campaign could also easily be the result of an unrelated event---e.g., a celebrity marriage or World Cup game~\cite{erhart_wikipedias_2019}. To account for these challenges, many researchers rely on some form of regression discontinuity design that focuses on changes between a short time period (e.g., two weeks) before and after an event (e.g., \cite{vincent_examining_2018,zhang_group_2011,moyer_determining_2015}). While powerful, this approach is limited to studying short-term effects and does not naturally lend itself to the task of nowcasting or forecasting.

In this paper, we explore the utility of Bayesian structural time series (BSTS) models for evaluating the impact of external events. BSTS models are designed to predict a given time series based on historical data, seasonality components, and additional control time series. They naturally incorporate uncertainty and the resulting forecast can then serve as a counterfactual---i.e. be compared against the actual time series following a given intervention to determine whether there is evidence that the intervention increased or decreased the magnitude of the time series. We test the BSTS model in two scenarios: the page preview roll-out in German and English Wikipedia as well as a video campaign in India designed to increase awareness about Hindi Wikipedia.

Our contributions are as follows:
\begin{itemize}
    \item \textbf{Page Previews}: using the roll-out of page previews on the German and English Wikipedia, we demonstrate that our BSTS model can effectively detect changes in page views given predictive control series.
    \item \textbf{Impact of video campaigns in India}: applying our BSTS model to online and TV awareness campaigns in India, we do not find evidence of increased page views as a result of the online or TV campaigns.
    \item \textbf{Correlations across languages and regions}: we evaluate the predictive power of page-view time-series between pairs of Wikipedia language editions and regions. We find evidence that page view trends are unique to a given country and language edition and that control series ideally originate from the same language edition followed by same country to be a useful predictor.

\end{itemize}

\section{Related Work}
In this work, we draw methods from the time series prediction literature and motivation from the literature that has sought to understand the impact of external events on Wikipedia activity.

\subsection{Time Series Modeling} \label{rw:timeseries}
The goal of the time series modeling that we employ in this paper is to understand whether a specific intervention significantly impacts a given metric for which we have temporal data---e.g., whether the roll-out of a new feature causes a change in daily traffic. There are many approaches to time series modeling that span from quite simple to much more complex in accordance with how many assumptions they make. Common to these models, however, is that their validity depends on the model being able to make direct comparisons between the time series prior to an intervention and the time series following the intervention~\cite{angrist_pischke_2009}. Threats to this validity come from a variety of sources that may affect time series independent of the intervention being studied: seasonality effects such as natural variation by day of week or month of year, unaccounted external events such as holidays or changes in the size of the underlying population. The likelihood that these events affect the time series increases as the time period being studied increases. A strong model, then, incorporates covariates that can control for seasonality, holidays, and other external factors that might affect the time series. A strong model also effectively represents its own uncertainty about predictions when there are insufficiently strong controls in place.

The core distinguishing features of approaches are 1) whether they include a control time series, and, 2) whether they directly compare metrics before and after the intervention or predict the time series after the intervention and compare this counterfactual prediction with the actual data. A control time series is a time series that is highly correlated with the ``treated'' times series, but, importantly, is known to not be affected by the intervention. The value of a control time series is that it helps to ensure that if seasonality or another event affects the treated time series, this effect is not conflated with the impact of the intervention because the effect should also be present in the control time series. The value of comparing the post-intervention time series with a counterfactual, as opposed to just the values from prior to the intervention, is greater flexibility to changing conditions. A model that produces the counterfactual can take into account more data about the time series prior to the intervention and, therefore, better account for shifts in covariates that might occur following the intervention. This is especially important when considering the long-term effects of an intervention. For these models, if the actual time series falls outside of the bounds of the counterfactual time series, this provides evidence that the intervention had a significant impact.

There are many considerations for how to build a robust time series model in order to produce the counterfactual predictions. Primarily, some models that use static regression to produce the counterfactual predictions falsely assume the data to be independent and identically distributed (i.i.d), which would result in an underestimation of the uncertainty~\cite{bertrand_duflo_mullainathan_2002}. Secondarily, to help avoid over-fitting, we need to choose appropriate control time series. Castle et al.~\cite{castle_qin_reed_2009} reviews and compares 21 methods for variable selection, including significance testing (e.g., forward and backward step-wise regression) and information criteria (e.g., AIC, BIC). Other popular model selection algorithms in time-series forecasting includes principal component and factor models, and penalized regression models (e.g., Lasso, ridge regression). However, these techniques force us to use a fixed set of selected variables, or do not account for the uncertainty in variable selection. Lastly, in order to gauge the uncertainty of the impact, we need to account for various sources of uncertainty in the model. Besides the uncertainty in variable selection and auto-correlation mentioned above, we also want to account for uncertainties in the historical relationships between treated and control time series, as well as uncertainties in seasonality and other components in the model.

\subsection{Impact of External Events on Wikipedia} \label{rw:wikipediaevents}

A number of papers have considered the challenge of establishing how an external event has affected dynamics within Wikipedia. Vincent et al.~\cite{vincent_examining_2018} and Moyer et al.~\cite{moyer_determining_2015} take what is known as a \emph{interrupted time series} (ITS) approach to examine how posts on Reddit that include links to Wikipedia articles affect page view traffic on Wikipedia. Both model a Reddit post with a Wikipedia link as a ``shock'' to that Wikipedia article and compare the mean number of page views in a short period of time before and after the post to determine whether there is a significant difference. Zhang and Zhu~\cite{zhang_group_2011} take a similar approach for the rate of contributions to Chinese Wikipedia from outside editors before and after a block on mainland China. Zhang and Zhu also seek to control for seasonal trends by examining the same time period in prior years. These approaches build on the assumption that there should be no difference in the expected page views between the pre- and post-intervention periods, and therefore any difference in page views can be causally tied to the Reddit post or block. As discussed above, these are assumptions that may hold true for short time-spans like one week, given that they do not happen to coincide with major holidays or events. This assumption is increasingly tenuous, however, as more long-term trends are considered.

Penney~\cite{penney_chilling_2016} also starts with an ITS model to understand the impact of the Edward Snowden revelations on page views to ``terrorism-related'' Wikipedia articles. Notably, because Penney examines a much longer time-period comprising 32 months, their analysis also includes a ``control time series'' that are security-related articles that are similar in content but less likely to be affected by the Snowden revelations. This approach is often referred to as difference-in-differences (DD) and is similar to how we construct the BSTS models considered in this work, but the BSTS models directly incorporate the concept of a control series as a core component of the models and captures the uncertainty of the relationship between the treated series and the control series. This makes for a much more explicit and robust means of controlling for additional external effects that may otherwise be conflated with the treatment under study.

\section{Methods}

We use a single time series model architecture, described below, and apply it to two different events. Each event involves an external event that led to a potential shift in page views. We describe each event alongside its results.

\subsection{Bayesian Structural Time Series Model}

In this work, we use Bayesian structural time series (BSTS) model \cite{scott_predicting_2013}. Per the components discussed in \S\ref{rw:timeseries}, these models can incorporate control covariates and time series, generate counterfactual predictions of page views for the post-intervention period assuming that the intervention did not take place, and naturally model their own uncertainty about these counterfactuals. We can then compare the counterfactual predictions and actual page views to quantify the causal impact of the intervention.

BSTS models combine three statistical methods into an integrated architecture \cite{NBERw19567}:
\begin{itemize}
    \item A structural time series model for trend and seasonality, estimated using Kalman filters;
    \item Spike and slab regression for variable selection;
    \item Bayesian model averaging for the final prediction.
\end{itemize}

\textbf{Structural Time Series Model}: Under different assumptions, a very large class of models can be expressed in the form of structural time series models, including all ARIMA models~\cite{harvey_1991}. This flexibility allows BSTS models to accommodate multiple sources of variations, including trends, seasonality, and latent evolutions of the treated series that cannot be explained by known trends or events. Specifically, a structural time series model (e.g. Eq.~\ref{locallinearmodel}) decomposes the time series into four components: a level ($\mu_{t}$), a local trend or slope ($\delta_{t}$), seasonal effects ($\tau_{t}$) and error terms. The model described here adds a regression component ($\beta^{T}\mathbf{x}_{t}$) to incorporate the control time series and other covariates. It is a stochastic generalization of the constant-trend regression model (e.g. $y_{t} = \mu +\delta t + \beta^{T}\mathbf{x}_{t} + \epsilon_{t}$), where the level $\mu_{t}$ and slope $\delta_{t}$ parameters each follow a random walk model instead of a constant. This allows for greater flexibility in the trends expressed within the model.

\begin{equation} \label{locallinearmodel}
\begin{aligned}
y_{t}  &= \mu _{t} + \tau _{t} + \beta^{T}\mathbf{x}_{t} + \epsilon _{t} , \epsilon _{t} \sim N(0, \sigma_\epsilon^{2})\\
\mu_{t} &= \mu_{t-1} + \delta_{t-1} + u_{t} , u _{t} \sim N(0, \sigma_u^{2}) \\
\delta_{t} &= \delta_{t-1} + v _{t}, v_{t} \sim N(0, \sigma_v^{2})\\
\tau _{t} &= -\sum_{s=1}^{S-1}\tau_{t-s} + w_{t} , w _{t} \sim N(0, \sigma_w^{2})
\end{aligned}
\end{equation}

\textbf{Spike and Slab}: There are often many potential control series but including them all would likely lead to over-fitting and very complex models. A spike-and-slab prior over coefficients~\cite{george_mcculloch_1997,madigand_raftery_1994} is designed to solve this challenge. The spike part controls the probability of whether a given variable would be chosen for the model---i.e. having a non-zero coefficient. The slab part shrinks the non-zero coefficients toward prior expectations which is often zero. Upon observing data, Bayes' theorem updates the inclusion probability of each coefficient. Then when sampling from the posterior distribution of a regression model, many of the simulated regression coefficients will be exactly zero~\cite{scoot_fitting_2017}.
 
\textbf{Bayesian Model Averaging}: To generate counterfactual predictions, the procedure uses the Markov chain Monte Carlo (MCMC) algorithm to draw samples from the parameter's posterior distribution and then combine that with the available data to yield a distribution of the counterfactual predictions. The model can then compute the difference between the actual values of a treated series in the post-intervention period and the distribution of counterfactual samples to yield an estimate of the distribution of the impact~\cite{brodersen_gallusser_koehler_remy_scott_2015}. Because the structural time series model, spike-and-slab regression and model averaging all have natural Bayesian interpretations, BSTS is able to account for various sources of uncertainties using MCMC. This allows us to gauge confidence in the magnitude of causal impact and estimate the posterior probability that the causal impact is non-existent.

\begin{figure*}[t]
    \centering
    \includegraphics[width=\linewidth]{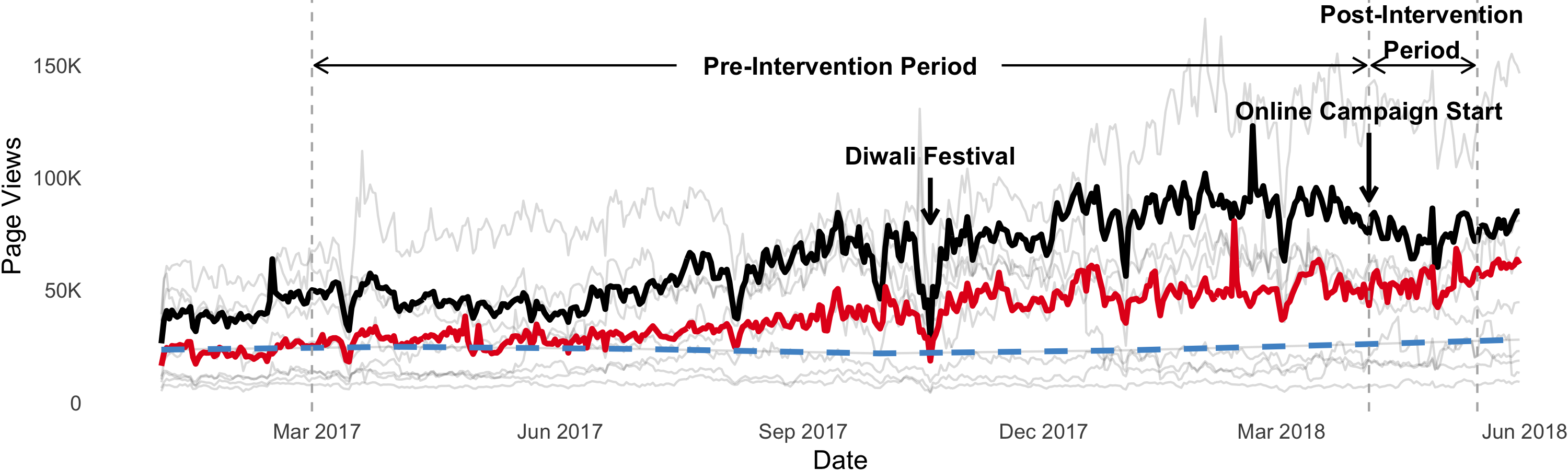}
    \caption{Hindi Wikipedia daily externally-referred page views from the top 10 states with the most page views in India. The black (top highlighted) series is the treated series: page views from Madhya Pradesh. The red (bottom highlighted) series is one of the most predictive control series according to the model in \S\ref{hivideo}: page views from the neighboring state of Rajasthan. The blue dashed line is a covariate: interpolated number of internet subscribers in thousands from Madhya Pradesh. The intervention under study (online video campaign) started on 3 April 2018.}
    \label{fig:model-param}
\end{figure*}

We use the \emph{BSTS}\footnote{https://cran.r-project.org/package=bsts} and \emph{CausalImpact}\footnote{https://cran.r-project.org/package=CausalImpact} R packages to fit the BSTS models. The following parameters comprise each BSTS model. Figure \ref{fig:model-param} illustrates the parameters of the model for Hindi online video campaign in \S\ref{hivideo}.
\begin{itemize}
    \item \textbf{Treated Time Series}: this is the time series under study---e.g., daily internally-referred page views to English Wikipedia over the period of several months.
    \item \textbf{Intervention}: an event that occurred on a specific date during the study period that is believed to have affected the treated time series---e.g., the introduction of a new feature onto Wikipedia that might change the daily number of page views. Model validation is conducted entirely on data prior to the intervention.
    \item \textbf{Pre-Intervention Period}: time period from the first data point of the treated series to the day before the intervention. For each model in this work, we explore four different pre-intervention period length using grid-search: 12 weeks, 18 weeks, 183 days and 400 days.
    \item \textbf{Post-Intervention Period}: time period following the intervention for which the impact is being estimated---e.g., daily page views for the six weeks following the roll-out of a new feature. For each model in this work, we set this to 6 weeks.
    \item \textbf{Covariates}: additional variables that help explain the treated time series---e.g., total population online in a country. This also includes the control series described below.
    \item \textbf{Control Series}: a time series that is predictive of the treated time series prior to the intervention, but that is not be impacted by the intervention---e.g., daily page views for a similar Wikipedia edition for which the feature was not rolled out. The authors of the \emph{CausalImpact} library we use for estimating the models suggest using 3-50 covariates.\footnote{https://stats.stackexchange.com/questions/162930/causalimpact-should-i-use-more-than-one-control/163554\#163554} Thus, for some models in this work where we have hundreds of control series---e.g., many combinations of different regions and language editions---we use correlation and dynamic time warping (DTW)~\cite{toni_computing_2009} algorithms with pre-intervention data to prescreen and trim the list of control series before feeding them into the BSTS model.
    \item \textbf{Seasonality}: weekly and seasonal trends, or holiday effects that did not get captured by the control series. For each model in this work, we include features for day-of-week and month. We also include major holidays for the regions under study as described in the Results section.
    \item \textbf{Trend Model}: the architecture for the model that predicts how the treated time series evolves. For each model in this work, we explore the following types of trend components using grid-search: local level, local linear, semi-local linear, and static intercept term.
\end{itemize}

\section{Results}

Below we describe the context for two events on which we tested our BSTS model and their results.

\subsection{Wikipedia Page Previews} \label{pvprollout}

Beginning in 2014, Wikipedia began exploring a new feature that would allow for page previews for the desktop version of the site. When a user moused over a link, a card would appear with an image and part of the first paragraph from the article that the link pointed to (see the Wikimedia Blog post\footnote{https://blog.wikimedia.org/2018/05/09/page-previews-documentation/} for more information and an example of a page preview on Wikimedia Commons\footnote{https://commons.wikimedia.org/w/index.php?curid=47213242}). This allows users to preview the article content without clicking on it (and thereby recording a page view). As a result of this ability to explore Wikipedia content without actually visiting the pages, it was expected that page views would actually drop with the roll-out of this feature.

In late 2017, the feature was finally rolled out to a proportion of anonymous users on German (de-wiki) and English (en-wiki) Wikipedia in a series of A/B tests. These tests were analyzed and it was determined that page preview feature led to a drop of approximately 4\% in page views across these Wikipedia communities.\footnote{https://www.mediawiki.org/wiki/Page\_Previews/2017-18\_A/B\_Tests} The full deployment of page previews to all anonymous users of these communities occurred on April 11 and 17 respectively. These A/B tests present us with an opportunity to explore the power of our BSTS models because they experimentally determined the expected effect on overall page views from the full deployment of page previews. Specifically, from the A/B tests starting on December 21 2017, in which 1.5\% and 4\% of anonymous users on English and German Wikipedias by default had access to page preview functionality respectively, we expect our BSTS model to detect:
\begin{itemize}
    \item \textbf{de-wiki}: a 3.0\% decrease in page views after April 11 2018.\footnote{https://phabricator.wikimedia.org/T191966}
    \item \textbf{en-wiki}: a 4.7\% decrease in page views after April 17 2018.\footnote{https://phabricator.wikimedia.org/T191101}
\end{itemize}

\subsubsection{Model Parameters}

For each model, we set the pre-intervention period to be 400 days and the post-intervention period to be 6 weeks. For English Wikipedia, this means that the time series starts on 13 March 2017 and includes daily page views data through 28 May 2018, with the intervention occurring on 17 April 2018. For German Wikipedia, this means that the time series starts on 7 March 2017 and includes daily page views data through 22 May 2018, with the intervention occurring on 11 April 2018. Alongside day-of-week and monthly seasonality, we also include the following holidays: Christmas and New Year's. For trend modeling, we choose a static intercept term for German Wikipedia---i.e. we expect the trend of the time series to be soaked up by the regression component, and a local level model for English Wikipedia---i.e. the trend will be predicted around the weighted average values of recent observations.

For the control series, we rely on the assumption that while page previews should impact the internally-referred page views---i.e. page views as a result of navigating from one Wikipedia page to another---there is no reason that the previews would impact externally-referred page views---i.e. page views that result from someone navigating from a search engine or other, non-Wikimedia website---or direct page views without referrer. Specifically, we select the daily internally-referred page views from en-wiki or de-wiki as the treated time series. For our control time series, we include the daily externally-referred page views and direct page views from the same Wikipedia language edition as the treated time series under study (i.e. en-wiki or de-wiki), as well as daily externally-referred page views from the other top-20 largest Wikipedia editions (e.g. Russian and Spanish Wikipedia).

\subsubsection{Results}

We find that our time series models for both English and German Wikipedia are quite accurate. The validation statistics associated with the model provide an indication of how effective the model was at predicting the pre-intervention time series. With 10-fold cross validation and prediction evaluated on 6 weeks of daily page views (from the end of the pre-intervention period), the holdout mean absolute percentage error (MAPE) of the English Wikipedia model is 2.54\%, and the holdout MAPE of the German Wikipedia model is 3.92\%.

\begin{figure*}[t]
    \centering
    \includegraphics[width=\linewidth]{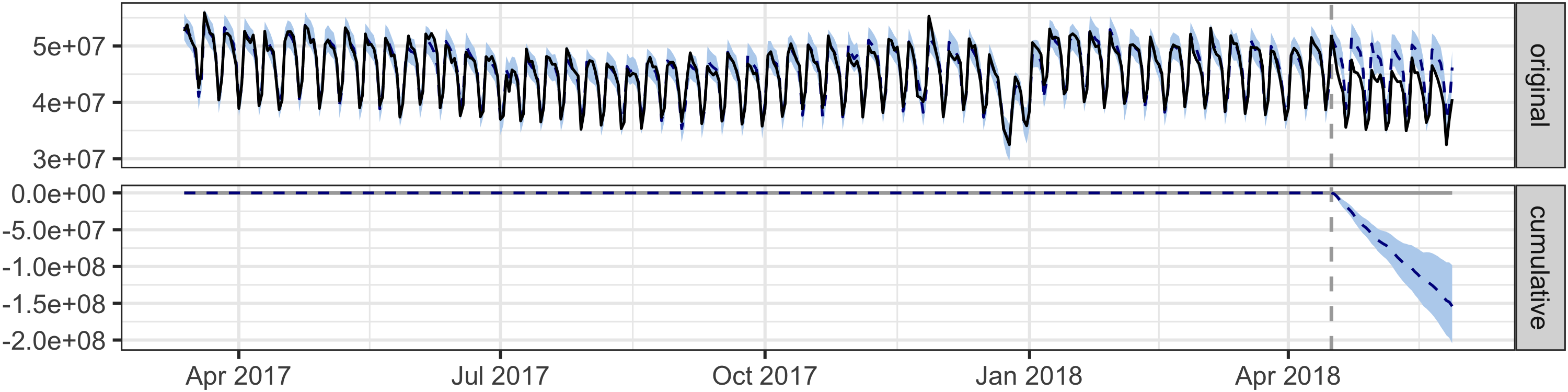}
    \caption{Impact of the page preview feature on 6 weeks of English Wikipedia internally-referred page views. Vertical dashed line represents the date of the roll-out (17 April 2018). Shaded areas indicate 95\% credible intervals. The first panel shows the data (black solid line) and counterfactual prediction (blue dashed line) for the post-intervention period. The second panel sums the difference between observed data and counterfactual predictions---i.e. point-wise causal effect as estimated by the model, resulting in a plot of the cumulative effect of the intervention. The figures of point-wise causal effect are removed in this paper for space consideration.}
    \label{fig:enwiki-pagepreviewgraphs}
\end{figure*}

Turning to the estimate of the causal impact of the page preview roll-out, the treated time series along with counterfactual estimates from model for en-wiki are shown in Figure~\ref{fig:enwiki-pagepreviewgraphs}. Recall that the early A/B tests indicated that there would be a 4.7\% decrease in page views for en-wiki. Our BSTS model, using the externally-referred and direct page views as control series, estimates a 3.0\% decrease and correctly determines that no impact---i.e. 0\% change---falls outside of the 95\% credible interval [1.9\%, 3.9\%], indicating that the roll-out resulted in a significant change in page views. The most predictive control series in this model is the search engine referred page views on English Wikipedia with an average standardized coefficient of 0.65---i.e. when search engine referred page views change 1 standard deviation, we expect to see internally-referred page views change 0.65 standard deviation, and the posterior inclusion probability---i.e. the probability of this coefficient being different from zero---is 100\%. 

We see analogous results for de-wiki: the BSTS model estimated a 2.6\% decrease in page views with a 95\% credible interval of [1.9\%, 3.4\%], in line with the 3.0\% decrease that had been determined via A/B testing. Similarly, the most predictive control series in this model is the search engine referred page views on German Wikipedia with an average standardized coefficient of 0.95 and a posterior inclusion probability of 100\%.

\subsection{Hindi Video Campaign} \label{hivideo}

In India, only 33\% of Hindi internet users have heard of Wikipedia~\cite{gill_how_2018} and, while there are 120,000 Wikipedia articles in Hindi, many people do not know that Hindi content exists. Meanwhile, internet access is growing 20\%+ per year across India\footnote{http://www.internetlivestats.com/internet-users/india/} and Hindi online content consumption is growing 94\% per year\footnote{https://economictimes.indiatimes.com/tech/internet/hindi-content-consumption-on-internet-growing-at-94-google/articleshow/48528347.cms}. In July 2017, the Wikimedia Foundation and the Hindi Wikimedians User Group began collaborating to reach ``New Readers'' in India through production and promotion of an online video.\footnote{https://meta.wikimedia.org/wiki/New\_Readers/Raising\_Awareness\_in\_India} The goal is to increase awareness and drive new usage of Wikipedia among Hindi speaking internet users.

To explain and promote Hindi Wikipedia (hi-wiki), the Wikimedia Foundation launched the video campaign on 3 April 2018. The Ektara\footnote{https://commons.wikimedia.org/wiki/File:Wikipedia\_-\_Ektara\_(English\_subtitles).webm} video was promoted on YouTube and Facebook targeting Hindi internet users in Madhya Pradesh, many of whom who had not heard of Wikipedia. The online promotion ran for three weeks and the video gathered 2.61 million views. This was followed by a second push over TV during a major Cricket event (on DD Sports during the Indian Premier League finals) on 27 May 2018 to the whole country, which reached 1.37 million viewers.\footnote{TV data was collected by Eurodata TV via BARC in India.}

\subsubsection{Model Parameters}

The pre-intervention period is 400 days and the post-intervention period is 6 weeks. Alongside day-of-week and monthly seasonality, we also include the following major Hindu holidays: Diwali, Raksha Bandhan, Holi, Dussehra, and New Year.\footnote{These Hindu festivals are picked because of their relative big impact on the treated time series. Their dates of each year are obtained from https://www.officeholidays.com/countries/india/index.php} Local level model and a static intercept term are chosen as the trend for online campaign model and TV campaign model respectively.

For the evaluation of the impact of the online campaign, we set the treated time series to be daily externally-referred page views to hi-wiki from the Indian state of Madhya Pradesh (as determined by IP geolocation) because the promotion of the online video campaign was targeted at Madhya Pradesh. We select the externally-referred page views because it is a good indicator of the general brand awareness. The time series starts on 27 February 2017 and includes daily page view data through 14 May 2018, where the intervention occurred on 3 April 2018. For the control series, we use daily hi-wiki page views, as well as page views to other popular Wikipedia language editions and Wikimedia projects,\footnote{We selected the top 10 Wikimedia projects in India with the most page views, and Wikipedia of major Indian languages spoken by more than 4\% of the population, according to 2011 census of India.} from the rest of India by states. We also included the daily number of internet subscribers in Madhya Pradesh as a covariate, which is linearly interpolated from a quarterly series reported by Telecom Regulatory Authority of India.\footnote{https://www.trai.gov.in/release-publication/reports/performance-indicators-reports}

For the evaluation of the impact of the TV campaign, we set the treated time series to be daily externally-referred hi-wiki page views from the entire country of India because there was no state-specific targeting of the campaign. The time series starts on 22 April 2017 and includes daily page view data through 7 July 2018, where the intervention occurred on 27 May 2018. For the control series, we use daily hi-wiki page views and page views to other popular Wikipedia language editions and Wikimedia projects from other countries\footnote{Countries that contribute more than 5\% of Hindi Wikipedia page views, countries whose official language is Hindi, and other nearby countries.}. Additionally, we included the daily number of internet subscribers in India as a covariate, which is linearly interpolated from a quarterly series reported by Telecom Regulatory Authority of India.

\subsubsection{Results}

\begin{figure*}[t]
    \centering
    \includegraphics[width=\linewidth]{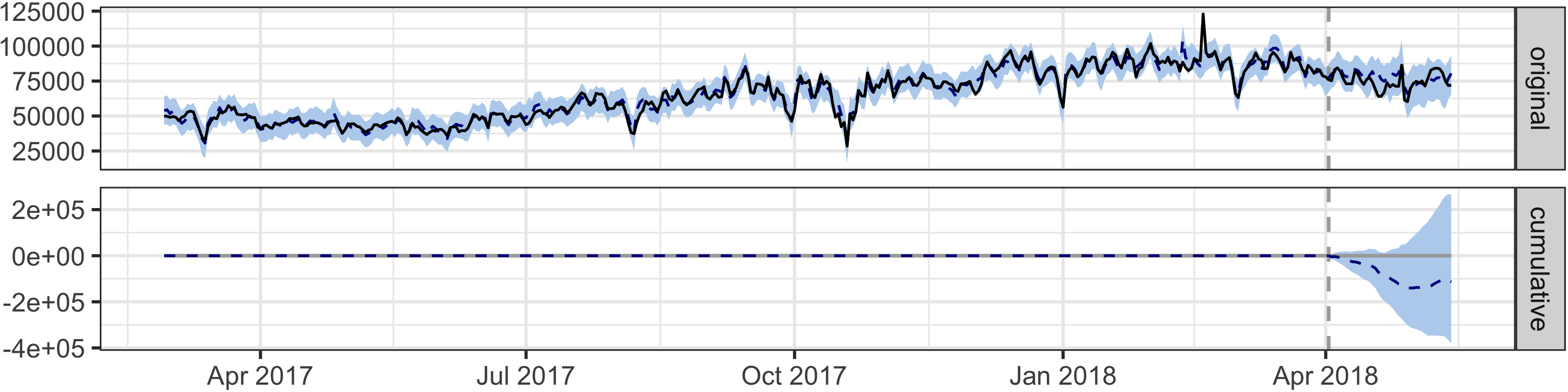}
    \caption{Impact of the online campaign in 6 weeks on Hindi Wikipedia externally-referred page views from Madhya Pradesh. Vertical dashed line represents the start date of the campaign 3 April 2018.}
    \label{fig:onlinecampaignimpact}
\end{figure*}

\begin{figure*}[t]
    \centering
    \includegraphics[width=\linewidth]{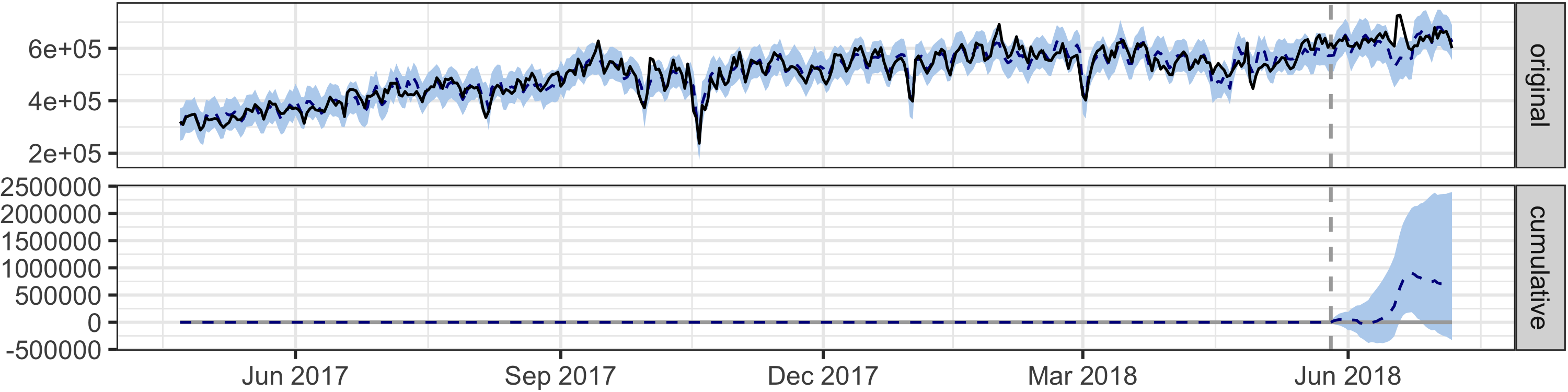}
    \caption{Impact of the TV campaign in 6 weeks on Hindi Wikipedia externally-referred page views from all of India. Vertical dashed line represents the date of the campaign 27 May 2018.}
    \label{fig:tvcampaignimpact}
\end{figure*}

For both the online and TV campaigns for Hindi Wikipedia, we do not detect a significant change in page views. As we discuss below and in \S\ref{choosecontrol}, this is likely a combination of factors: low impact and imprecise control series. The results from the BSTS models for the online campaign are in Figure~\ref{fig:onlinecampaignimpact} and TV campaign in Figure~\ref{fig:tvcampaignimpact}.

First we examine the results for the online video campaign that was targeted at the state of Madhya Pradesh. As before, we performed a 10-fold cross validation with the pre-intervention time series and predict 6-week's daily page views in each fold. The average holdout MAPE is 7.6\%. 

As Figure~\ref{fig:onlinecampaignimpact} indicates, no significant impact on page views was detected following the intervention. While there does appear to be a downward trend in page views, zero change is still within the 95\% credible interval. Page views to hi-wiki from the states of Rajasthan and Chhattisgarh, both of which border Madhya Pradesh, are the most predictive control series in the model, with average standardized coefficients of 0.23 and 0.13 respectively. The posterior probabilities that their coefficients are different from zero are greater than 95\%.

Next we turn to the results for the country-wide TV campaign. The average holdout MAPE from the cross validation is 10.2\%. Figure~\ref{fig:tvcampaignimpact} shows the results from the BSTS model. As the graphs indicate, no significant impact on page views was detected in the first 3 weeks following the intervention. There was a bump in the 4th week after the campaign, but it is most likely to be the result of an unknown event. Overall, we did not detect significant impact in 6 weeks. The number of internet subscribers in India, hi-wiki page views from the United States, Bengali Wikipedia page views from Bangladesh and English Wiktionary pageviews from Nepal are the most predictive control series in the model, with an average standardized coefficient of 0.84, 0.35, 0.14 and 0.16 respectively. The posterior probabilities that their coefficients are different from zero are greater than 95\%.

We did not include the page views to other Wikipedia language editions that also were geolocated to India---e.g., page views to en-wiki from India---in the set of control series because it may violate the independence assumption. Most people in India are multilingual, so if our brand awareness was affected by the campaign, the impact would likely be revealed on page views of other Wikipedia language editions from the target region as well. After seeing the relatively high MAPE (10.2\%) of the TV campaign model, we tried to include the page views to other Wikipedia language editions and other Wikimedia projects from India into the model to see if they help. The average holdout MAPE decreased to 8.5\%, but that model also does not detect significant impact.

\section{Discussion}

\subsection{Choosing a Control Time Series} \label{choosecontrol}

\begin{table*}[t]
\centering
\begin{tabular}{|c|c|c|c|c|}
\hline
                 & \textbf{Treated Series (lang; region)}            & \textbf{Control Series (lang; region)}                   & \textbf{Avg MAPE} & \textbf{\begin{tabular}[c]{@{}c@{}}Avg MAPE\\ (No Control Series)\end{tabular}} \\ \hline
\textbf{Model 1} & hi-wiki page views; Madhya Pradesh & hi-wiki page views; other states of India & 7.54\%                    & 11.54\%                                                                                 \\ \hline
\textbf{Model 2} & hi-wiki page views; Madhya Pradesh & other wikis page views; Madhya Pradesh    & 9.31\%                    & 11.54\%                                                                                 \\ \hline
\textbf{Model 3} & hi-wiki page views; all of India          & hi-wiki page views; other countries       & 7.22\%                    & 7.92\%                                                                                  \\ \hline
\textbf{Model 4} & hi-wiki page views; all of India          & other wikis page views; all of India             & 9.14\%                    & 7.92\%                                                                                  \\ \hline

\end{tabular}
\caption{Comparison of the predictive power of four sets of control time series for the Hindi Wikipedia campaign, of which the treated series and the control series either share the same geographic region or the same Wikipedia language edition.}
\label{table:controlcomparison}
\end{table*}

A predictive control series is one of the most important aspects of a BSTS model and also the part of the model that is often most difficult to choose. From the page preview roll-out analysis (\S\ref{pvprollout}), we see that the BSTS model, with a well-chosen control time series, can effectively estimate the impact of a given external event. In that case, externally-referred and internally-referred page views for the same project are highly correlated, but only internally-referred page views were believed to be impacted by the page preview roll-out. For both German and English Wikipedia, the estimated causal impact was slightly conservative---i.e. lower in magnitude than expected based on the A/B tests---but still quite close to the expected impact. We had less success selecting an effective control series for the Hindi video campaign. Furthermore, the fact that the online campaign model for Madhya Pradesh had less error than the TV campaign model for all of India raises questions about how factors like geography or language edition affect the predictive power of a control series for Wikipedia.

To better understand the power of different types of control series, we tested four additional control time series models for the Hindi analysis. All are trained on 400 days of daily externally-referred page views and evaluated via 10-fold cross validation on 6 weeks of page views prior to the intervention (3 April 2018). All models include day-of-week and monthly seasonality, holiday effects, local level trends, and a set of control time series but no further covariates. The control series for each model and its respective validation error are shown in Table~\ref{table:controlcomparison}.

We see that the control time series that are from the same language edition as the treated time series (Models 1 and 3) have a consistently lower error than the models that are from different language edition but the same geographic region (Models 2 and 4). This indicates that language edition plays a more important role than geographic region in the page view trends on Wikipedia. Comparing Model 1 and 3 where the control and treated series are from the same language edition, page views between states within the same country are more predictive of each other (adding these controls into the model decrease the MAPE from 11.54\% to 7.54\%) than page views between different countries (adding them into the model only decrease the MAPE from 7.92\% to 7.22\%), which indicates that while language appears to be most important, country borders are still a highly salient aspect of page view trends on Wikipedia. Further research would be needed to understand how these effects play out in other language communities and the inter-relatedness of different countries and language pairs.

\section{Future Work and Limitations}
While this work has a number of limitations, as we lay out below, we believe it lays the groundwork for exploring more standardized methods of predicting trends such as page views on Wikipedia with the goal of understanding the effect of external events. Limitations for this work largely relate to temporal evolution of impact, data pre-processing, prior distributions of parameters in BSTS, and the need for additional experiments.

In this work, we focus only on the cumulative effect by the end of the post-intervention period---its existence and magnitude---without discussing the temporal evolution of an impact. In practice, how an effect evolves over time, especially its onset and decay structure, is often a valuable question as well. The point-wise effect from BSTS reflects the temporal evolution and future implementation should consider analyzing this result.

Small volume Wikipedia editions such as hi-wiki are more sensitive to undetected bot behavior, which can cause anomalies in page-view data. Anomalies in the prediction or post-intervention period would increase the error rate of validation, or the model might detect an impact that is unrelated to the known intervention. When the number of control series is very large, removing outliers manually is not feasible and thus requires a robust algorithm to detect and adjust outliers while preserving those known ``outliers'' such as holiday effects. It is possible that further pre-processing would also provide benefits---e.g., including more holidays, removing seasonal patterns in predictors before fitting the model, more extensive grid search for parameters like the length of pre-intervention period.

We were expecting that spike-and-slab prior in the BSTS would prevent over-fitting by forcing the coefficients of poor predictors to zero, so we would at least have predictions not worse than that of a model which only contains the historical information of the treated series itself. Contrary to this expectation, Model 4 from Table~\ref{table:controlcomparison} (predicting hi-wiki page views in India) shows that including control series from other Wikipedia language editions and other Wikimedia projects within India actually added noise to the prediction. To solve this problem, we can further tune the hyper-parameter that controls the expected model size---the expected number of coefficients that are different from zero---so that when most of the predictors do not have enough predictive power, we can lower the expected model size and force more coefficients to be zero (we set the expected model size to be 10\% of total number of controls in the model but not greater than 5 in this work).

Finally, future work should continue to explore these models in more contexts. This would hopefully provide guidance for how to select control time series---e.g., which pairs of regions and language editions (or even other Wikimedia projects) are predictive, what is the best way to split Wikipedias into control and treatment articles, how to take advantage of more granular information as with the internal/external referrer information. This would also hopefully provide guidance for how to set priors in BSTS models---e.g., a prior likelihood of relationships between treatment and control time series or prior standard deviation of the Gaussian random walk of the trend models (conservatively, we use a non-informative prior for the former and 0.01 for the latter). In future analyses, we can increase the prior inclusion probabilities for control series that are likely to be correlated with the treated series, and increase the prior standard deviation for the trend models if we believe the volatility of residuals would be large after regressing out known predictors.

%
\begin{acks}
We would like to thank Tilman Bayer, Mikhail Popov, Leila Zia, Kate Zimmerman and Jon Katz for guidance and support as well as our anonymous reviewers for their feedback.
\end{acks}

%
\bibliographystyle{ACM-Reference-Format}
\balance
\bibliography{bibliography}


\begin{thebibliography}{24}


\ifx \showCODEN    \undefined \def \showCODEN     #1{\unskip}     \fi
\ifx \showDOI      \undefined \def \showDOI       #1{#1}\fi
\ifx \showISBNx    \undefined \def \showISBNx     #1{\unskip}     \fi
\ifx \showISBNxiii \undefined \def \showISBNxiii  #1{\unskip}     \fi
\ifx \showISSN     \undefined \def \showISSN      #1{\unskip}     \fi
\ifx \showLCCN     \undefined \def \showLCCN      #1{\unskip}     \fi
\ifx \shownote     \undefined \def \shownote      #1{#1}          \fi
\ifx \showarticletitle \undefined \def \showarticletitle #1{#1}   \fi
\ifx \showURL      \undefined \def \showURL       {\relax}        \fi
\providecommand\bibfield[2]{#2}
\providecommand\bibinfo[2]{#2}
\providecommand\natexlab[1]{#1}
\providecommand\showeprint[2][]{arXiv:#2}

\bibitem[\protect\citeauthoryear{Alexa}{Alexa}{2018}]%
        {alexa_wikipedia.org_2018}
\bibfield{author}{\bibinfo{person}{Alexa}.} \bibinfo{year}{2018}\natexlab{}.
\newblock \bibinfo{booktitle}{\emph{wikipedia.org {Traffic} {Statistics}}}.
\newblock \bibinfo{type}{{T}echnical {R}eport}. \bibinfo{institution}{Alexa}.
\newblock
\urldef\tempurl%
\url{https://www.alexa.com/siteinfo/wikipedia.org}
\showURL{%
\tempurl}


\bibitem[\protect\citeauthoryear{Angrist and J{\"o}rn-Steffen}{Angrist and
  J{\"o}rn-Steffen}{2009}]%
        {angrist_pischke_2009}
\bibfield{author}{\bibinfo{person}{Joshua~D. Angrist} {and}
  \bibinfo{person}{Pischke J{\"o}rn-Steffen}.} \bibinfo{year}{2009}\natexlab{}.
\newblock \bibinfo{booktitle}{\emph{Mostly Harmless Econometrics: An
  Empiricists Companion}}.
\newblock \bibinfo{publisher}{Princeton University Press}.
\newblock


\bibitem[\protect\citeauthoryear{Bertrand, Duflo, and Mullainathan}{Bertrand
  et~al\mbox{.}}{2002}]%
        {bertrand_duflo_mullainathan_2002}
\bibfield{author}{\bibinfo{person}{Marianne Bertrand}, \bibinfo{person}{Esther
  Duflo}, {and} \bibinfo{person}{Sendhil Mullainathan}.}
  \bibinfo{year}{2002}\natexlab{}.
\newblock \showarticletitle{How Much Should We Trust Differences-in-Differences
  Estimates?}
\newblock  (\bibinfo{year}{2002}).
\newblock
\urldef\tempurl%
\url{https://doi.org/10.3386/w8841}
\showDOI{\tempurl}


\bibitem[\protect\citeauthoryear{Brodersen, Gallusser, Koehler, Remy, and
  Scott}{Brodersen et~al\mbox{.}}{2015}]%
        {brodersen_gallusser_koehler_remy_scott_2015}
\bibfield{author}{\bibinfo{person}{Kay~H. Brodersen}, \bibinfo{person}{Fabian
  Gallusser}, \bibinfo{person}{Jim Koehler}, \bibinfo{person}{Nicolas Remy},
  {and} \bibinfo{person}{Steven~L. Scott}.} \bibinfo{year}{2015}\natexlab{}.
\newblock \showarticletitle{Inferring causal impact using Bayesian structural
  time-series models}.
\newblock \bibinfo{journal}{\emph{The Annals of Applied Statistics}}
  \bibinfo{volume}{9}, \bibinfo{number}{1} (\bibinfo{year}{2015}),
  \bibinfo{pages}{247--274}.
\newblock
\urldef\tempurl%
\url{https://doi.org/10.1214/14-aoas788}
\showDOI{\tempurl}


\bibitem[\protect\citeauthoryear{Castle, Qin, and Reed}{Castle
  et~al\mbox{.}}{2009}]%
        {castle_qin_reed_2009}
\bibfield{author}{\bibinfo{person}{Jennifer~L. Castle},
  \bibinfo{person}{Xiaochuan Qin}, {and} \bibinfo{person}{W.~Robert Reed}.}
  \bibinfo{year}{2009}\natexlab{}.
\newblock \bibinfo{title}{How To Pick The Best Regression Equation: A Review
  And Comparison Of Model Selection Algorithms, by Jennifer L. Castle;
  Xiaochuan Qin; W. Robert Reed}.
\newblock
\newblock
\urldef\tempurl%
\url{https://ideas.repec.org/p/cbt/econwp/09-13.html}
\showURL{%
\tempurl}


\bibitem[\protect\citeauthoryear{Clark, Faris, and Heacock~Jones}{Clark
  et~al\mbox{.}}{2017}]%
        {clark_analyzing_2017}
\bibfield{author}{\bibinfo{person}{Justin Clark}, \bibinfo{person}{Robert
  Faris}, {and} \bibinfo{person}{Rebekah Heacock~Jones}.}
  \bibinfo{year}{2017}\natexlab{}.
\newblock \bibinfo{booktitle}{\emph{Analyzing {Accessibility} of {Wikipedia}
  {Projects} {Around} the {World}}}.
\newblock \bibinfo{type}{Berkman {Klein} {Center} for {Internet} \& {Society}
  {Research} {Publication}}.
\newblock


\bibitem[\protect\citeauthoryear{Erhart}{Erhart}{2019}]%
        {erhart_wikipedias_2019}
\bibfield{author}{\bibinfo{person}{Ed Erhart}.}
  \bibinfo{year}{2019}\natexlab{}.
\newblock \bibinfo{title}{Wikipedia's most-popular articles of 2018 show that
  pop culture rules over us all}.
\newblock
\newblock
\urldef\tempurl%
\url{https://wikimediafoundation.org/2019/01/02/wikipedias-most-popular-articles-of-2018-show-that-pop-culture-rules-over-us-all/}
\showURL{%
\tempurl}


\bibitem[\protect\citeauthoryear{George and McCulloch}{George and
  McCulloch}{1997}]%
        {george_mcculloch_1997}
\bibfield{author}{\bibinfo{person}{E.~I. George} {and} \bibinfo{person}{R.~E.
  McCulloch}.} \bibinfo{year}{1997}\natexlab{}.
\newblock \showarticletitle{Approaches for Bayesian variable selection}.
\newblock \bibinfo{journal}{\emph{Statistica Sinica}}  \bibinfo{volume}{7}
  (\bibinfo{year}{1997}), \bibinfo{pages}{339--374}.
\newblock


\bibitem[\protect\citeauthoryear{Gill and McCune}{Gill and McCune}{2018}]%
        {gill_how_2018}
\bibfield{author}{\bibinfo{person}{Satdeep Gill} {and} \bibinfo{person}{Zack
  McCune}.} \bibinfo{year}{2018}\natexlab{}.
\newblock \bibinfo{title}{How we're building awareness of {Wikipedia} in
  {India}}.
\newblock
\newblock
\urldef\tempurl%
\url{https://blog.wikimedia.org/2018/04/03/building-awareness-wikipedia-india/}
\showURL{%
\tempurl}


\bibitem[\protect\citeauthoryear{Harvey}{Harvey}{1991}]%
        {harvey_1991}
\bibfield{author}{\bibinfo{person}{Andrew~C. Harvey}.}
  \bibinfo{year}{1991}\natexlab{}.
\newblock \bibinfo{booktitle}{\emph{Forecasting, structural time series models
  and the Kalman filter}}.
\newblock \bibinfo{publisher}{Cambridge Univ. Press}.
\newblock


\bibitem[\protect\citeauthoryear{Madigan and Raftery}{Madigan and
  Raftery}{1994}]%
        {madigand_raftery_1994}
\bibfield{author}{\bibinfo{person}{D. Madigan} {and} \bibinfo{person}{A.~E.
  Raftery}.} \bibinfo{year}{1994}\natexlab{}.
\newblock \showarticletitle{Model selection and accounting for model
  uncertainty in graphical models using Occam's window}.
\newblock \bibinfo{journal}{\emph{J. Amer. Statist. Assoc.}}
  \bibinfo{volume}{89} (\bibinfo{year}{1994}), \bibinfo{pages}{1535--1546}.
\newblock


\bibitem[\protect\citeauthoryear{Mesty{\'a}n, Yasseri, and
  Kert{\'e}sz}{Mesty{\'a}n et~al\mbox{.}}{2013}]%
        {mestyan_early_2013}
\bibfield{author}{\bibinfo{person}{M{\'a}rton Mesty{\'a}n},
  \bibinfo{person}{Taha Yasseri}, {and} \bibinfo{person}{J{\'a}nos
  Kert{\'e}sz}.} \bibinfo{year}{2013}\natexlab{}.
\newblock \showarticletitle{Early prediction of movie box office success based
  on {Wikipedia} activity big data}.
\newblock \bibinfo{journal}{\emph{PloS one}} \bibinfo{volume}{8},
  \bibinfo{number}{8} (\bibinfo{year}{2013}), \bibinfo{pages}{e71226}.
\newblock


\bibitem[\protect\citeauthoryear{Moat, Curme, Avakian, Kenett, Stanley, and
  Preis}{Moat et~al\mbox{.}}{2013}]%
        {moat_quantifying_2013}
\bibfield{author}{\bibinfo{person}{Helen~Susannah Moat},
  \bibinfo{person}{Chester Curme}, \bibinfo{person}{Adam Avakian},
  \bibinfo{person}{Dror~Y Kenett}, \bibinfo{person}{H~Eugene Stanley}, {and}
  \bibinfo{person}{Tobias Preis}.} \bibinfo{year}{2013}\natexlab{}.
\newblock \showarticletitle{Quantifying {Wikipedia} usage patterns before stock
  market moves}.
\newblock \bibinfo{journal}{\emph{Scientific reports}}  \bibinfo{volume}{3}
  (\bibinfo{year}{2013}), \bibinfo{pages}{1801}.
\newblock


\bibitem[\protect\citeauthoryear{Moyer, Carson, Dye, Carson, and
  Goldbaum}{Moyer et~al\mbox{.}}{2015}]%
        {moyer_determining_2015}
\bibfield{author}{\bibinfo{person}{Daniel~Cheng Moyer},
  \bibinfo{person}{Samuel~L Carson}, \bibinfo{person}{Thayne~Keegan Dye},
  \bibinfo{person}{Richard~T Carson}, {and} \bibinfo{person}{David Goldbaum}.}
  \bibinfo{year}{2015}\natexlab{}.
\newblock \showarticletitle{Determining the {Influence} of {Reddit} {Posts} on
  {Wikipedia} {Pageviews}}. In \bibinfo{booktitle}{\emph{Ninth {International}
  {AAAI} {Conference} on {Web} and {Social} {Media}}}.
\newblock


\bibitem[\protect\citeauthoryear{Penney}{Penney}{2016}]%
        {penney_chilling_2016}
\bibfield{author}{\bibinfo{person}{Jonathon~W Penney}.}
  \bibinfo{year}{2016}\natexlab{}.
\newblock \showarticletitle{Chilling effects: {Online} surveillance and
  {Wikipedia} use}.
\newblock \bibinfo{journal}{\emph{Berkeley Tech. LJ}}  \bibinfo{volume}{31}
  (\bibinfo{year}{2016}), \bibinfo{pages}{117}.
\newblock


\bibitem[\protect\citeauthoryear{Priedhorsky, Osthus, Daughton, Moran,
  Generous, Fairchild, Deshpande, and Del~Valle}{Priedhorsky
  et~al\mbox{.}}{2017}]%
        {priedhorsky_measuring_2017}
\bibfield{author}{\bibinfo{person}{Reid Priedhorsky}, \bibinfo{person}{Dave
  Osthus}, \bibinfo{person}{Ashlynn~R Daughton}, \bibinfo{person}{Kelly~R
  Moran}, \bibinfo{person}{Nicholas Generous}, \bibinfo{person}{Geoffrey
  Fairchild}, \bibinfo{person}{Alina Deshpande}, {and} \bibinfo{person}{Sara~Y
  Del~Valle}.} \bibinfo{year}{2017}\natexlab{}.
\newblock \showarticletitle{Measuring global disease with {Wikipedia}:
  {Success}, failure, and a research agenda}. In
  \bibinfo{booktitle}{\emph{Proceedings of the 2017 {ACM} {Conference} on
  {Computer} {Supported} {Cooperative} {Work} and {Social} {Computing}}}.
  \bibinfo{publisher}{ACM}, \bibinfo{pages}{1812--1834}.
\newblock


\bibitem[\protect\citeauthoryear{Scott}{Scott}{2017}]%
        {scoot_fitting_2017}
\bibfield{author}{\bibinfo{person}{Steven~L Scott}.}
  \bibinfo{year}{2017}\natexlab{}.
\newblock \bibinfo{title}{Fitting Bayesian structural time series with the bsts
  R package}.
\newblock
\newblock
\urldef\tempurl%
\url{http://www.unofficialgoogledatascience.com/2017/07/fitting-bayesian-structural-time-series.html}
\showURL{%
\tempurl}


\bibitem[\protect\citeauthoryear{Scott and Varian}{Scott and Varian}{2013a}]%
        {NBERw19567}
\bibfield{author}{\bibinfo{person}{Steven~L Scott} {and} \bibinfo{person}{Hal~R
  Varian}.} \bibinfo{year}{2013}\natexlab{a}.
\newblock \bibinfo{booktitle}{\emph{Bayesian Variable Selection for Nowcasting
  Economic Time Series}}.
\newblock \bibinfo{type}{Working Paper} 19567. \bibinfo{institution}{National
  Bureau of Economic Research}.
\newblock
\urldef\tempurl%
\url{https://doi.org/10.3386/w19567}
\showDOI{\tempurl}


\bibitem[\protect\citeauthoryear{Scott and Varian}{Scott and Varian}{2013b}]%
        {scott_predicting_2013}
\bibfield{author}{\bibinfo{person}{Steven~L Scott} {and} \bibinfo{person}{Hal~R
  Varian}.} \bibinfo{year}{2013}\natexlab{b}.
\newblock \showarticletitle{Predicting the present with bayesian structural
  time series}.
\newblock \bibinfo{journal}{\emph{Available at SSRN 2304426}}
  (\bibinfo{year}{2013}).
\newblock


\bibitem[\protect\citeauthoryear{Singer, Lemmerich, West, Zia, Wulczyn,
  Strohmaier, and Leskovec}{Singer et~al\mbox{.}}{2017}]%
        {singer_why_2017}
\bibfield{author}{\bibinfo{person}{Philipp Singer}, \bibinfo{person}{Florian
  Lemmerich}, \bibinfo{person}{Robert West}, \bibinfo{person}{Leila Zia},
  \bibinfo{person}{Ellery Wulczyn}, \bibinfo{person}{Markus Strohmaier}, {and}
  \bibinfo{person}{Jure Leskovec}.} \bibinfo{year}{2017}\natexlab{}.
\newblock \showarticletitle{Why {We} {Read} {Wikipedia}}. In
  \bibinfo{booktitle}{\emph{Proceedings of the 26th {International}
  {Conference} on {World} {Wide} {Web}}}. \bibinfo{publisher}{International
  World Wide Web Conferences Steering Committee}, \bibinfo{pages}{1591--1600}.
\newblock


\bibitem[\protect\citeauthoryear{ten Thij, Volkovich, Laniado, and
  Kaltenbrunner}{ten Thij et~al\mbox{.}}{2012}]%
        {ten_thij_modeling_2012}
\bibfield{author}{\bibinfo{person}{Marijn ten Thij}, \bibinfo{person}{Yana
  Volkovich}, \bibinfo{person}{David Laniado}, {and} \bibinfo{person}{Andreas
  Kaltenbrunner}.} \bibinfo{year}{2012}\natexlab{}.
\newblock \showarticletitle{Modeling and predicting page-view dynamics on
  {Wikipedia}}.
\newblock \bibinfo{journal}{\emph{CoRR abs/1212.5943}} (\bibinfo{year}{2012}).
\newblock


\bibitem[\protect\citeauthoryear{Toni}{Toni}{2009}]%
        {toni_computing_2009}
\bibfield{author}{\bibinfo{person}{Giorgino Toni}.}
  \bibinfo{year}{2009}\natexlab{}.
\newblock \showarticletitle{Computing and {Visualizing} {Dynamic} {Time}
  {Warping} {Alignments} in {R}: {The} dtw {Package}}.
\newblock \bibinfo{journal}{\emph{Journal of Statistical Software}}
  \bibinfo{volume}{31} (\bibinfo{year}{2009}).
\newblock
\urldef\tempurl%
\url{https://doi.org/10.18637/jss.v031.i07}
\showDOI{\tempurl}


\bibitem[\protect\citeauthoryear{Vincent, Johnson, and Hecht}{Vincent
  et~al\mbox{.}}{2018}]%
        {vincent_examining_2018}
\bibfield{author}{\bibinfo{person}{Nicholas Vincent}, \bibinfo{person}{Isaac
  Johnson}, {and} \bibinfo{person}{Brent Hecht}.}
  \bibinfo{year}{2018}\natexlab{}.
\newblock \showarticletitle{Examining {Wikipedia} {With} a {Broader} {Lens}:
  {Quantifying} the {Value} of {Wikipedia}'s {Relationships} with {Other}
  {Large}-{Scale} {Online} {Communities}}. In
  \bibinfo{booktitle}{\emph{Proceedings of the 2018 {CHI} {Conference} on
  {Human} {Factors} in {Computing} {Systems}}}. \bibinfo{publisher}{ACM},
  \bibinfo{pages}{566}.
\newblock


\bibitem[\protect\citeauthoryear{Zhang and Zhu}{Zhang and Zhu}{2011}]%
        {zhang_group_2011}
\bibfield{author}{\bibinfo{person}{Xiaoquan~Michael Zhang} {and}
  \bibinfo{person}{Feng Zhu}.} \bibinfo{year}{2011}\natexlab{}.
\newblock \showarticletitle{Group size and incentives to contribute: {A}
  natural experiment at {Chinese} {Wikipedia}}.
\newblock \bibinfo{journal}{\emph{American Economic Review}}
  \bibinfo{volume}{101}, \bibinfo{number}{4} (\bibinfo{year}{2011}),
  \bibinfo{pages}{1601--15}.
\newblock


\end{thebibliography}

\end{document}